\documentclass[conference]{IEEEtran}
\IEEEoverridecommandlockouts

\usepackage{amsmath}
\usepackage{amssymb}
\usepackage{amsthm}
\usepackage{blindtext}
\usepackage[noadjust]{cite}
\usepackage{epsf}
\usepackage{epsfig}
\usepackage{etoolbox}
\usepackage{float}
\usepackage{graphicx}
\usepackage{hhline}
\usepackage{latexsym}
\usepackage{multirow}
\usepackage{psfrag}
\usepackage[usenames,dvipsnames]{pstricks}
\usepackage{pst-plot}
\usepackage{stfloats}
\usepackage{subfig}
\usepackage{tcolorbox}
\usepackage{balance}
\usepackage{algorithm}
\usepackage{algpseudocode}

\algrenewcommand\algorithmicrequire{\textbf{Input:}}
\algrenewcommand\algorithmicensure{\textbf{Output:}}

\graphicspath{ {figures/} }

\newcommand{\ceil}[1]{\lceil{#1}\rceil}

\newcommand{\iid}{i.\@i.\@d.\ }

\theoremstyle{definition}\newtheorem{lemma}{Lemma}
\theoremstyle{definition}
\theoremstyle{definition}
\theoremstyle{definition}

\newtheorem{remark}[lemma]{Remark}

\newcommand\shortintertext[1]{%
\ifvmode\else\\\@empty\fi
\noalign{%
\penalty0%
\vbox{\mathstrut}%
\penalty10000%
\vskip-\baselineskip
\penalty10000%
\vbox to 0pt{%
\normalbaselines
\ifdim\linewidth=\columnwidth
\else
\parshape\@ne
\@totalleftmargin\linewidth
\fi
\vss
\noindent#1\par}%
\penalty10000%
\vskip-\baselineskip}%
\penalty10000}

\newtheorem{thm}{Theorem}
\newtheorem{definition}{Definition}

\newtheorem{fact}[thm]{Fact}

\begin{document}
\title{
Syndrome-based Fusion Rules in Heterogeneous Distributed Quickest Change Detection
{\thanks {This work was supported by the National Science and Technology Council of Taiwan under grants MOST 111-2221-E-A49-069-MY3 and MOST 111-2811-E-A49-519-MY3.}}
}

\author{\vspace{-0.5cm}\\ Wen-Hsuan Li and Yu-Chih Huang\\
Institute of Communications Engineering\\
National Yang Ming Chiao Tung University,  Taiwan\\
E-mail:\{vincent, jerryhuang\}@nycu.edu.tw
}
%%%%%%
% \author{%
%   \IEEEauthorblockN{Anonymous Authors}
%    \IEEEauthorblockA{%
%      Please do NOT provide authors' names and affiliations\\
%      in the paper submitted for review, but keep this placeholder.\\
%      ISIT23 follows a \textbf{double-blind reviewing policy}.}
% }

\maketitle
\thispagestyle{empty}
\pagestyle{empty}

\begin{abstract}

In this paper, the heterogeneous distributed quickest change detection (HetDQCD) with 1-bit non-anonymous feedback is studied. The concept of syndromes is introduced and the family of syndrome-based fusion rules is proposed, which encompasses all deterministic fusion rules as special cases. Through the Hasse diagram of syndromes, upper and lower bounds on the second-order performance of expected detection delay as a function of average run length to false alarm are provided. An interesting instance, the weighted voting rule previously proposed in our prior work, is then revisited, for which an efficient pruning method for breadth-first search in the Hasse diagram is proposed to analyze the performance. This in turn assists in the design of the weight threshold in the weighted voting rule. Simulation results corroborate that our analysis is instrumental in identifying a proper design for the weighted voting rule, demonstrating consistent superiority over both the anonymous voting rule and the group selection rule in HetDQCD.

\end{abstract}

\section{Introduction} \label{Introduction}
% Survey on classical QCD
Quickest change detection (QCD) focuses on quickly detecting an unknown change time point so that the false alarm is below a tolerance constraint. Several early works \cite{ref:Page54, ref:Lorden71, ref:Moustakides86} have developed fundamental results for this classical problem. The well-known solution for the criterion in \cite{ref:Lorden71}, called the cumulative sum (CUSUM) test, was first proposed by Page in \cite{ref:Page54}, which was proved to be asymptotically and exactly optimal by Lorden \cite{ref:Lorden71} and Moustakides \cite{ref:Moustakides86}, respectively. Many applications of this problem can be found in statistical analysis and signal processing \cite{ref:Lai08, ref:Lakhina04, ref:Nizam16, ref:Zhang23}.

% Survey on (homogeneous) distributed QCD and Byzantine distributed QCD (with iot or indust4.0)
Since the rise of applications on distribution systems such as the internet of things (IoT) and distributed learning/computing, how to efficiently detect abrupt changes in a distributed system has been a crucial issue, motivating the study of distributed QCD (DQCD) \cite{ref:Liyan21_survey}. 
%is a promising solution \cite{ref:Liyan21_survey}, seeing one example on the smart grid \cite{ref:Chen16}. 
In DQCD problem, the task of detection is commonly carried out by a fusion center through distributed sensors that provide feedback based on their observations/decisions via band-limited links. Several important works of performance analysis on DQCD with limited feedback were made by Mei \cite{ref:Mei05} and Banerjee and Fellouris \cite{ref:Banerjee16}. However, in the DQCD problem, it is also challenging with the discrepancy between the quality of feedback information. For example, the DQCD problem with feedback discrepancy caused by Byzantine attacks have been studied in \cite{ref:Fellouris18} and \cite{ref:Huang21}. Another less extreme example is the heterogeneous distributed QCD (HetDQCD) problem \cite{ref:Sun22}, where it was assumed that every sensor is honest, but the observations of each sensor are drawn from an independent but possibly nonidentical distribution to model heterogeneity. Sun et al. demonstrated in the same study that the mixture CUSUM procedure is the optimal anonymous rule when observations are directly available at the fusion center. In \cite{ref:Sun23}, the fusion center is assumed to receive unlabeled data samples, and a generalized mixture CUSUM algorithm is proved by Sun et al. to be second-order asymptotically optimal. 

In our previous work \cite{ref:ISIT23}, we addressed the HetDQCD problem with {\it 1-bit feedback} and focused on performance loss due to anonymity. With second-order performance analysis, we proposed the parameters for anonymous voting rule. Moreover, two non-anonymous fusion rules, namely the group selection rule and the weighted voting rule, were proposed.
Simulation results therein showed that the knowledge of the identity of feedback may indeed be turned into performance enhancement. 
%By comparing the simulation result to non-anonymous weighted voting rule in heuristic setting, we observe the gain from feedback identity. 

Extending the prior work in \cite{ref:ISIT23}, we further investigate the non-anonymous setting of HetDQCD. Specifically, we introduce the concept of syndromes and turn every deterministic fusion rule for HetDQCD into a syndrome-based fusion rule. A special type of syndromes called critical syndromes, is identified.
%provide a generalized expression based on feedback syndromes for the simultaneous fusion rule. 
With the concept of critical syndromes on the Hasse diagram, the performance bounds of such a syndrome-based fusion rule can be obtained by observing the sufficient and necessary conditions to trigger an alarm at the fusion center. 
%In particular, these conditions are highly correlated with a specific collection of syndromes called critical syndromes. 
Under the framework of syndrome-based fusion rules, the low-complexity yet very effective weighted voting rule previously proposed in \cite{ref:ISIT23} is revisited. To analyze the performance, a pruning algorithm for breadth-first search in the corresponding Hasse diagram is proposed to find its dominant critical syndromes efficiently. This analysis is then exploited to design the weight threshold for the weighted voting rule. 
%While the general fusion rule is difficult to capture overall syndromes, we revisit weighted voting rule proposed in \cite{ref:ISIT23}, which is more trackable since the selecting of syndromes only depends on the weighted sum. With the concept of syndrome, we derive the second-order approximation of the weighted voting rule, and the suggested threshold can be obtained by solving some integer optimization problems. Notably, these optimization problems can be solved efficiently based on the structure of the Hasse diagram.
Simulation results show that the weighted voting rule with suggested threshold outperforms both the anonymous voting rule and group selection rule. %, no matter whether the sensor network is highly heterogeneous or not. The performance gap between the weighted and anonymous voting rules adequately characterizes the performance loss in anonymity.

The remainder of this paper is organized as follows. Section \ref{ProblemFormulation} introduces the system model and the problem formulation. Section \ref{SyndFusionRule} introduces the proposed syndrome-based fusion rules and the performance analysis. Section \ref{WeightedFusionRule} revisits the weighted voting rule under the framework of syndrome-based fusion rules. Computer simulations to verify these analytic results are given in Section \ref{SimulationResults}, followed by the conclusion in Section \ref{Conclusions}.

\section{Background} \label{ProblemFormulation}

% In this section, we introduce the network model, the local detection rule adopted at each sensor, and two fusion rules considered in this paper. The problem studied in this paper is to carefully analyze these schemes and provide efficient detection rule for HetDQCD with 1-bit feedback based on the analysis.

We first present the heterogeneous sensor network model and the local detection rule used by each sensor. An early result of the simultaneous fusion rule in \cite{ref:ISIT23} is then shown below.

\subsection{Problem formulation}
% \begin{figure}[t]
% 	\centering
% 	\includegraphics[scale=0.35]{fig_SensorNetworks.png}
% 	\caption{The heterogeneous wireless sensor network. \vspace{-0.2cm}\\}
% 	\label{network}
% \end{figure}

We consider a heterogeneous sensor network composed of $N$ sensors and one fusion center. With the help of sensors, the goal of the fusion center is to detect an abrupt change that occurs at time $\nu$. Depending on the heterogeneous types, these $N$ sensors can be divided into $L$ groups, namely $\mathcal{G}_1, \ldots, \mathcal{G}_L$. Denote $\mathcal{G}$ be the union of all groups. In group $l\in[L]$, we represent the index of the $k$th sensor by pair $(l,k)$, and let $N_l$ be the number of sensors in this group. At time $t$, the sensor $(l,k)$ will observe $X_{t}^{l,k}$, which is drawn independently following
\begin{equation}
X_{t}^{l,k} \sim \left\{
                 \begin{array}{ll}
                   f_l, & \hbox{$t\leq\nu$,} \\
                   g_l, & \hbox{$t>\nu$,}
                 \end{array}
               \right.
\end{equation}
in which $f_l$ and $g_l$ are the pre-change and post-change distributions, respectively, with the positive Kullback-Leibler divergence (KLD) $\mathcal{I}_{l}>0$. Without loss of generality, we assume that $0 < \mathcal{I}_{1} \leq \mathcal{I}_{2} \cdots \leq \mathcal{I}_{L} < \infty$.
% We consider the heterogeneous sensor network as illustrated in Fig. \ref{network}, where $N$ sensors assist the fusion center in detecting an abrupt change occurring at time $\nu$. There are total $L$ heterogeneous types and thereby the $N$ sensors can be divided into $L$ groups, namely $\mathcal{G}_1, \ldots, \mathcal{G}_L$, according to their types. Let $N_l$ be the number of sensors of type $l\in[L]$ and we use the pair $(l,k)$ to denote the index of sensor $k\in[N_l]$ in group $l\in[L]$. The sensor $(l,k)$ will observe $X_{t}^{l,k}$ at time $t$ that is drawn independently according to
% \begin{equation}
% X_{t}^{l,k} \sim \left\{
%                  \begin{array}{ll}
%                    f_l, & \hbox{$t\leq\nu$,} \\
%                    g_l, & \hbox{$t>\nu$,}
%                  \end{array}
%                \right.
% \end{equation}
% where $f_l$ and $g_l$ are the pre-changed and post-changed distributions, respectively, and they have the positive Kullback-Leibler divergence (KLD) $\mathcal{I}_{l}>0$. %\textcolor{blue}{Based on the standard assumption of the asymptotic analysis from Lorden's framework \cite{ref:Lorden71}, the KLD for each group is assumed to be positive and finite.}

At time $t$, each sensor has the ability to transmit a 1-bit noiseless message to the fusion center. Feedback is said to be non-anonymous if the sensor index $(l,k)$ of the feedback message is known; otherwise, it is said to be anonymous.
%to fusion center are assumed to be either anonymous or with full identity.} 
Subsequently, the fusion center employs a fusion rule $\rho$ to make a decision based on the received feedback. In particular, let $\rho$ be an $\mathcal{F}_{t}$-stopping time, where 
\begin{align}
       \mathcal{F}_{t}\equiv \sigma \left(X_{s}^{l,k};s \in [t],k \in [N_{l}],l \in [L]\right),
\end{align}
is the $\sigma$-algebra from all possible observations of all sensors up to time $t$. 
%{\blue Notice that anonymous feedback will restrict fusion rules that work without feedback identity.} 
Let $P_{\nu}$ represent the underlying probability measure of the event that occurs at time $\nu$. Specifically, in the absence of the event, we assign $\nu = \infty$, and denote the corresponding underlying probability measure as $P_{\infty}$. Two metrics, namely the average run length (ARL) to false alarms and the worst-case expected detection delay (EDD), to evaluate the performance of $\rho$ are defined as $\textrm{ARL}(\rho) = {E_{\infty}[\rho]}$ and
\begin{align}
    \textrm{EDD}(\rho) &=\sup_{\nu\geq 0}\mathrm{ess}\sup E_{\nu}[(\rho-\nu)^{+}|\mathcal{F}_{\nu}],
\end{align}
respectively, in which $E_{\nu}$ is the expectation under $P_{\nu}$. By \cite[Lemma 3]{ref:Fellouris18}, the above definition can be further simplified to 
\begin{equation}
    \mathrm{EDD}(\rho)=E_0[\rho] \textrm{,}
\end{equation}
as long as the fusion rule $\rho$ only depends on the local CUSUM statistics. Following Lorden's framework \cite{ref:Lorden71}, the goal of HetDQCD is to minimize $\textrm{EDD}(\rho)$ such that $\textrm{ARL}(\rho)>\gamma$.
% At time $t$, each sensor can communicate an 1-bit message with the fusion center noiselessly. A fusion rule $\rho$ is then adopted by the fusion center to form a decision according to the feedbacks. Specifically, let $\rho$ be an $\mathcal{F}_{t}$-stopping time, in which 
% \begin{align}\mathcal{F}_{t}\equiv \sigma(X_{s}^{l,k};s \in [t],k \in [N_{l}],l \in [L]),
% \end{align}
% is the $\sigma$-algebra from all possible observations of all sensors up to time $t$. We denote by $P_{\nu}$ the underlying probability measure that the event happens at time $\nu$. In particular, if no event occurs, we set $\nu = \infty$ and denote by $P_{\infty}$ the underlying probability measure. Two metrics are defined to measure the performance of $\rho$, namely the average run length (ARL) to false alarm and the worst-case expected detection delay (EDD), as $\textrm{ARL}(\rho) = {E_{\infty}[\rho]}$, and
% \begin{align}
%     \textrm{EDD}(\rho) &=\sup_{\nu\geq 0}\mathrm{ess}\sup E_{\nu}[(\rho-\nu)^{+}|\mathcal{F}_{\nu}],
% \end{align}
% %$\textrm{ARL}(\rho) = {E_{\infty}[\rho]}$ and $\textrm{EDD}(\rho) =\sup_{\nu\geq 0}\mathrm{ess}\sup E_{\nu}[(\rho-\nu)^{+}|\mathcal{F}_{\nu}]$, 
% respectively, where $E_{\nu}$ is the expectation under $P_{\nu}$. Following Lorden's setting \cite{ref:Lorden71}, the problem of QCD is to minimize $\textrm{EDD}(\rho)$ subject to a constraint $\textrm{ARL}(\rho)>\gamma$.        

\subsection{Local CUSUM and voting rules}\label{subsec:review_voting}
Similar to most works in the literature \cite{ref:Mei05, ref:Banerjee16, ref:Fellouris18}, each sensor makes a local decision according to its CUSUM statistic. Specifically, upon observing $X_t^{l,k}$ at time $t$, sensor $(l,k)$ recursively computes the CUSUM statistic $W_{t}^{l,k}$ based on the log-likelihood ratio (LLR) $Z_{t}^{l,k}\equiv \log(f_l(X_t^{k.l})/g_l(X_t^{l,k}))$ by
\begin{align}
    W_{t}^{l,k} = \max{ \left\{0,W_{t-1}^{l,k} \right\}} + Z_{t}^{l,k} \textrm{, }\quad W_{0}^{l,k} \triangleq 0 \textrm{.}
\end{align}
Limited by the 1-bit bandwidth constraint, the sensor makes the local decision and sends 1-bit feedback on whether $W^{l,k}_t>h_l$ at time $t$, in which $h_l$ is the threshold for sensors in type $l\in[L]$. The fusion center then makes a global decision on whether there is a change or not. 
% {\blue With anonymous feedbacks, one efficient fusion rule is the anonymous $M$ voting rule \cite{ref:Banerjee16}: a global decision is reached for the initial occurrence of receiving at least $M$ alarms simultaneously. Denote $\mathbf{h} =[h_{1},...,h_{L}]^{T}$, a formal definition is described by}
% \begin{align}
%      \rho_{M}( \mathbf{h} ) = \inf \left\{t: \rvert \{(l,k):W_{t}^{l,k}>h_l \} \rvert \geq M \right\}\textrm{.}
% \end{align}
% {\blue (don't call it $\rho_M$; it may be confused with $\rho_{\Omega}$.)}
Under non-anonymous feedback, one naive strategy discussed in our previous work \cite{ref:ISIT23} only takes feedback from a subset of sensors into account and ignore feedback from all other sensors.
%aims to avoid triggering the false alarm via proper selecting informative sensors. 
Specifically, given the predetermined set $\mathcal{D}$, the anonymous voting rule $M$ within $\mathcal{D}$ is given by
\begin{equation}
    \hspace{-8pt}\textrm{T}_{M}( \mathbf{h},\mathcal{D} ) = \inf \left\{t: \left| \{(l,k)\in\mathcal{D}:W_{t}^{l,k}>h_l \} \right| \geq M \right\}\textrm{.} \label{def:MVotingWithinD}
\end{equation}
That is, a global decision is reached for the initial occurrence of receiving at least $M$ alarms within $\mathcal{D}$ simultaneously. It will reduce to the anonymous $M$ voting rule \cite{ref:Banerjee16} by setting $\mathcal{D} = \mathcal{G}$ and to the group selection rule by setting $\mathcal{D}=\mathcal{G}_L$.

In what follows, we summarize the theoretical analysis of the performance of $\textrm{T}_{M}(\mathbf{h},\mathcal{D})$\footnote{We present the result with $\mathcal{D}=\mathcal{G}$, which can be straightforwardly adapted to any subset $\mathcal{D}$ by replacing the network parameters.} in \cite{ref:ISIT23}, which will be crucial to the analysis in this work. 
%For theoretical analysis, our previous work \cite{ref:ISIT23} dedicated on the anonymous fusion rule. 
%Proposed by \cite{ref:ISIT23}, we assume that the threshold $h_l$ is proportional to the KLD, i.e. $h_l=\mathcal{I}_{l}h$ for some $h>0$, $l \in [L]$. %Theorem~\ref{thm:anonymous_voting_rule} summarizes the analytic results (after some adjustment) for the anonymous $M$ voting rule, which is the foundation for analyzing non-anonymous fusion rules. 

\begin{thm} \label{thm:anonymous_voting_rule}
Let $\mathbf{h} \mathnormal=[\mathcal{I}_{1}h,\cdots,\mathcal{I}_{L}h]^{T}$. Assume that
\begin{align}
    \sigma_{l}^{2} \equiv E_{0}\left[ \left( Z^{l,k}_{t}-\mathcal{I}_{l} \right)^{2} \right] < \infty \textrm{ for all } l,k.
\end{align}
And define $\lambda:[N] \rightarrow [L]$ by
\begin{align}
    \lambda (m) = l \textrm{~~if~~}\sum_{j=1}^{l-1}{N_{j}} \leq m \leq \sum_{j=1}^{l}{N_{j}} \textrm{.} \label{def:lambda}
\end{align}
Then as $h \rightarrow \infty$,
\begin{align}
    &E_{\infty} [\textrm{T}_{M} (\mathbf{h},\mathcal{G})] \sim \mathbf{\Theta}(1) e^{\sum_{m=1}^{M}{\mathcal{I}^{(\lambda (m))}}h} \textrm{,} \label{equ:FAR_mvote} \\
    &E_{0}[\textrm{T}_{M}(\mathbf{h},\mathcal{G})] =  h + \xi _{M}\sqrt{h}(1+o(1)),
    \label{equ:2nd_ADD_mvote}
\end{align}
\footnote{Only $\leq$ in \eqref{equ:2nd_ADD_mvote} was proved in \cite{ref:ISIT23}. We extend it to equality in Appendix~\ref{pf:2nd_ADD_mvote}.}where $\xi_{M}$ is the expected value of $M$-th order statistics of independent (but not necessarily identical) Gaussian random variables $G_{l,k} \sim N(0,\sigma_{l}^{2} / \mathcal{I}_{l}^2)$ for $1 \leq k \leq N_{l}$ and $1 \leq l \leq L$. In particular, if $h = h_{\gamma}$ leads to $E_{\infty} [\textrm{T}_{M} (\mathbf{h},\mathcal{G})] = \gamma$, then as $\gamma \rightarrow \infty$,
\begin{align}
    h_{\gamma} \sim \frac {\log \gamma} {\sum_{m=1}^{M}{\mathcal{I}_{\lambda (m)}}}. \label{equ:thres_mvote}
\end{align} 
\end{thm}

\section{Syndrome-based Fusion Rule} \label{SyndFusionRule}

\begin{figure}[tb]
	\centering
	\includegraphics[width=3.2in]{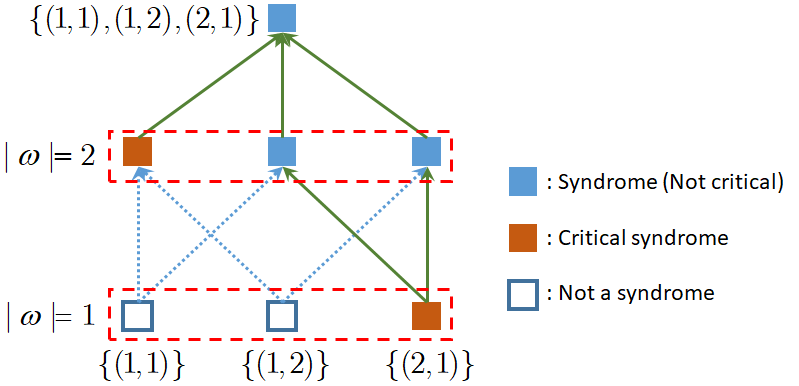}
	\caption{An example of general fusion rule. Any positive feedback from $(2,1)$ or the consensus decision from group $1$ will trigger the fusion center.}
	\label{fig:hasse_example}
\end{figure}

In this section, we propose a family of fusion rules, called syndrome-based fusion rules, for non-anonymous HetDQCD.

We first note that since the feedbacks are binary, every simultaneous deterministic fusion rule can actually be expressed as a combination of Boolean operations that map all possible received feedbacks to trigger or not. We call the pattern of received feedbacks a syndrome that triggers the fusion center. The general syndrome-based fusion rule can be expressed as the following:
\begin{align}
\rho_{\Omega}( \mathbf{h} ) = \inf\{t: \exists \omega \in \Omega \text{ s.t. } \mathbf{1}\{(l,k) \in \omega :W_{t}^{l,k}>h_l \} = 1 \} \textrm{,}    
\end{align}
where $\omega$'s are the syndromes that can trigger fusion center, and $\Omega$ is the collection of all the syndromes. Apparently, this general fusion rule contains {\it every} deterministic non-anonymous fusion rule. Moreover, it subsumes the anonymous $M$ voting rule as a special case, by setting
\begin{align}
\Omega = \{\omega:\rvert \omega \rvert \geq M \} \text{.}
\end{align}

Since the best decision made by the origin network is always more reliable than its subnetwork, we make the following assumptions of $\Omega$ by nature: 
\begin{itemize}
\item $\emptyset \notin \Omega$ and $\cup_{l=1}^{L}\mathcal{G}_l \in \Omega$.
\item If $\omega_{1} \in \Omega$ and $\omega_{1} \subseteq \omega_{2}$, then $\omega_{2} \in \Omega$.
\end{itemize}
Thus, these assumptions induce a partial order of set inclusion between syndromes. More specifically, given a general fusion rule $\Omega$, the syndromes form a connected subgraph on the Hasse diagram \cite{ref:Hasse} of power sets. An example is shown in Fig. \ref{fig:hasse_example}. With directed and acyclic properties, we can identify syndromes with no input edges, reflecting the necessary conditions to trigger the fusion center. A formal definition of these syndromes is given below.

\begin{definition} \label{def:critical_synd}
Let $\Omega$ be a collection of syndromes. A syndrome $\omega \in \Omega$ is called a critical syndrome if any subset of $\omega$ is not in $\Omega$. Denote by $\partial \Omega$ the collection of critical syndromes in $\Omega$.
\end{definition}

The analysis of critical syndromes is important since they can immediately induce a performance bound of $\Omega$ fusion rule by comparing to the voting rule within certain subnetworks:

\begin{thm} \label{thm:omega_fusion_rule}
For a general fusion rule $\Omega$, denote
\begin{align*}
m_{\Omega} = \mathop{\min \limits_{\omega \in \partial \Omega}}~{\rvert \omega \rvert} \textrm{ and } 
\mathcal{I}_{\Omega} = \mathop{\min \limits_{\omega \in \partial \Omega}}~{\sum_{(l,k) \in \omega}{\mathcal{I}_l}} \textrm{.}
\end{align*}
Suppose that $h = h_{\gamma,\Omega}$ leads to $E_{\infty} [\rho_{\Omega} (\mathbf{h})] = \gamma$. As $\gamma \rightarrow \infty$, for any pair $(M_{\Omega},D_{\Omega})$ such that the syndromes of $\textrm{T}_{M_{\Omega}}(\mathbf{h},\mathcal{D}_{\Omega})$ lie in $\Omega$, it follows that
\begin{align}
    h_{\gamma,\Omega}& \sim
    \frac {\log \gamma} {\mathcal{I}_{\Omega}}\textrm{,} \label{equ:thres_omega} \\
    h_{\gamma,\Omega}& + \xi _{m_{\Omega}}\sqrt{h_{\gamma,\Omega}}(1+o(1))
    \leq E_{0}[\rho_{\Omega}(\mathbf{h})] \nonumber \\ 
    &\leq h_{\gamma,\Omega} + \xi_{M_{\Omega},D_{\Omega}}\sqrt{h_{\gamma,\Omega}}(1+o(1))\textrm{,} \label{equ:EDD_omega}
\end{align}
in which $\xi_{M_{\Omega},\mathcal{D}_{\Omega}}$ is the expected value of $M_{\Omega}$-th order statistics of independent (but not necessarily identical) Gaussian random variables $G_{l,k} \sim N(0,\sigma_{l}^{2} / \mathcal{I}_{l}^2)$ for $(l,k) \in \mathcal{D}_{\Omega}$.
\end{thm}

\begin{IEEEproof}
See Appendix~\ref{pf:omega_fusion_rule}
\end{IEEEproof}

Although Theorem \ref{thm:omega_fusion_rule} provides a performance bound to the general case of simultaneous fusion, it is very difficult to design a proper fusion rule according to this theorem, due to the large number of combinations to choose syndromes. Even for a specific collection of syndromes $\Omega$, solving optimization problems corresponding to $m_{\Omega}$ and $\mathcal{I}_{\Omega}$ in Theorem \ref{thm:omega_fusion_rule} is NP-hard, not to mention the variation of $(M_{\Omega},\mathcal{D}_{\Omega})$. In what follows, we return to examine the weighted $M$ voting rule, which leads us to an explicit design of the parameters $m_{\Omega}$, $M_{\Omega}$ and $\mathcal{D}_{\Omega}$.

% advantage of weighted fusion rule: choosing synd by total weight summation=> less complexity than look-up table from general fusion rule.

\section{Weighted $M$ Voting Rule} \label{WeightedFusionRule}
For a trackable analysis, we carry on the concept of syndrome in Section \ref{SyndFusionRule} to analyze the weighted $M$ voting rule previously proposed in \cite{ref:ISIT23}, which is an intuitive (yet efficient) way to apply the knowledge of the feedback identity to the voting rule. It admits extremely low fusion complexity as the fusion center only needs to compute the weighted sum of the feedback.

Recall that for the anonymous $M$ voting rule, the fusion center will treat each feedback equally. For a non-anonymous scenario, once the fusion center obtains the knowledge for KLD of each group, the confidence of each feedback can be weighted, which induces the following weighted sum version of the fusion rule:
\begin{align}
    \bar{\rho}_{M}( \mathbf{h},\boldsymbol{\alpha} ) =
    \inf \left\{t: \sum_{l,k} {\alpha_{l,k}\mathbf{1}{\{W_{t}^{l,k}>h_l\}} } \geq M \right\}
    \textrm{,}
\end{align}
in which $\alpha_{l,k}$ is the weight for the $(l,k)$ sensor, and $\boldsymbol{\alpha} =[\alpha_{1,1},...,\alpha_{L,N_{L}}]^{T}$.
% special case: 0-1 weight => group selection
%             : proportional to  KLD
The performance of the weighted voting rule $M$ can be maintained by properly choosing the weight and the trigger threshold $M$. An extreme way of weight design is the anonymous voting rule within $\mathcal{D}$ reviewed in Section~\ref{subsec:review_voting} by setting
\begin{equation}
\alpha_{l,k} = \left\{
                 \begin{array}{ll}
                   1, & \hbox{$(l,k)\in \mathcal{D}$,} \\
                   0, & \hbox{otherwise.}
                 \end{array}
               \right.
\end{equation}
%which is to keep the most informative group only by setting
%\begin{equation}
%\alpha_{l,k} = \left\{
%                 \begin{array}{ll}
%                   1, & \hbox{$l = L$,} \\
%                   0, & \hbox{otherwise.}
%                 \end{array}
%               \right.
%\end{equation}
%In this case $\bar{\rho}_{M}( \mathbf{h},\boldsymbol{\alpha} ) = \textrm{T}_{M}( \mathbf{h},\mathcal{G}_{L})$. The group selection scheme can reduce the false alarms caused by those non-informative groups. 

To better utilize feedback from all the sensors, in \cite{ref:ISIT23} we provided a heuristic setting for the weight to be proportional to KLD and set $\alpha_{l,k} = 1$ for $1 \leq k \leq N_{L}$, or equivalently, we set
\begin{align}
\alpha_{l,k} = \mathcal{I}_{l}/\mathcal{I}_{L} \textrm{ for } 1 \leq k \leq N_{l} \textrm{ and } 1 \leq l \leq L \textrm{.}
\end{align}
% ==================================================
% Performance analysis: decision what M is the best => knapsack problem
%We follow the concept of syndrome and try to deal with the natural question: how these new fusion rules perform. For the group selection scheme, it reduces to the homogeneous subnetwork formed by the top informative group, whose analysis is similar to the anonymous case. 
Under these weights, we establish the connection between the weighted $M$ voting rule and the syndrome-based fusion. More specifically, we show that the performance of the weighted $M$ voting rule strongly depends on the critical syndromes determined by $M$. These critical syndromes obey the following conditions:
\begin{align}
\left\{
                 \begin{array}{ll}
                 \sum_{(l,k) \in \omega} {\mathcal{I}_{l}} \geq M\mathcal{I}_{L} \hbox{,} \\
                 \sum_{(l,k) \in \omega'} {\mathcal{I}_{l}} < M\mathcal{I}_{L} \hbox{, for all $\omega' \subsetneq \omega$.}
                 \end{array}
               \right. \label{equ:condition_csynd_weighted}
\end{align}
With a slight abuse of notation, let $\partial{M}$ be the collection of the above critical syndromes. With this connection, we can then apply Theorem~\ref{thm:omega_fusion_rule} to the following result.

\begin{thm} \label{thm:weighted_voting_rule}
For the weighted $M$ voting rule, denote $\bar{m} = \mathop{\min \limits_{\omega \in \partial{M}}}~{\rvert \omega \rvert}$ and $\bar{M} = \mathop{\max \limits_{\omega \in \partial{M}}}~{\rvert \omega \rvert}$. Suppose that $h = \bar{h}_{\gamma,M}$ leads to $E_{\infty} [\bar{\rho}_{M}( \mathbf{h},\boldsymbol{\alpha} )] = \gamma$. As $\gamma \rightarrow \infty$, it follows that
\begin{align}
    \bar{h}_{\gamma,M}& \leq 
    \frac {\log \gamma} {M \mathcal{I}_{L}}(1+o(1))\textrm{,} \label{equ:thres_weighted_M}\\
    \bar{h}_{\gamma,M}& + \xi_{\bar{m}}\sqrt{\bar{h}_{\gamma,M}}(1+o(1)) \leq E_{0}[\bar{\rho}_{M}( \mathbf{h},\boldsymbol{\alpha} )] \nonumber\\ 
    &\leq
    \bar{h}_{\gamma,M} + \xi_{\bar{M},\bar{\mathcal{D}}}\sqrt{\bar{h}_{\gamma,M}}(1+o(1))\textrm{,} \label{equ:EDD_weighted_M}
\end{align}
in which $\bar{\mathcal{D}}$ is a set of $(l,k)$ such that for any $(l_1,k_1)$,$\cdots$,$(l_{\bar{M}},k_{\bar{M}})$ in $\bar{\mathcal{D}}$, ${\sum_{m=1}^{\bar{M}}{\mathcal{I}_{l_m}}} \geq M\mathcal{I}_{L}$.
\end{thm}
\begin{IEEEproof}
    See Appendix~\ref{pf:weighted_voting_rule}.
\end{IEEEproof}

% Discussions
% 1. range of m_bar and M_bar => the effect on D_bar, accuracy of bounds and the relation to heterogeneity
\begin{remark}
The range between $\bar{m}$ and $\bar{M}$, and the maximal achieved set $\bar{\mathcal{D}}$ give a sketch of the selection of syndrome and the tightness of the bounds in \eqref{equ:EDD_weighted_M}. For example, if $\bar{m} = \bar{M}$ and $\bar{\mathcal{D}}$ contains all sensors, then the weighted $M$ voting rule is just the anonymous $\bar{M}$ voting rule. This can be proved by comparing the critical syndromes of the two fusion rules. For a general case, by setting
\begin{align}
\omega^{*} = arg \mathop{\min \limits_{\omega \in \partial{M},\rvert \omega \rvert = \bar{M}}}~{\sum_{(l,k) \in \omega}{\mathcal{I}_l}} \textrm{,}
\end{align}
the set $\bar{\mathcal{D}}$ can be chosen by
\begin{align}
\bar{\mathcal{D}} = \omega^{*} \cup \left\{ (l,k) \rvert l \geq \mathop{\max \limits_{(l',k') \in \omega^{*}}~l' } \right\} \textrm{.} \label{equ:D_bar}
\end{align}
\end{remark}
% 2. Algorithm for m_bar, M_bar and D_bar
\begin{remark}
With the asymptotic results \eqref{equ:thres_weighted_M} and \eqref{equ:EDD_weighted_M} in Theorem~\ref{thm:weighted_voting_rule}, we have
\begin{align}
E_{0}[\bar{\rho}_{M}( \mathbf{h},\boldsymbol{\alpha} )] \leq
    \left( \frac {\log \gamma} {M \mathcal{I}_{L}} + \xi_{\bar{M},\bar{\mathcal{D}}}\sqrt{\frac {\log \gamma} {M \mathcal{I}_{L}}} \right)(1+o(1)) \textrm{.} \label{equ:EDDARL_weighted_M}
\end{align}
A proper choice of $M$ can be obtained by minimizing the second-order approximation in \eqref{equ:EDDARL_weighted_M}. Following \eqref{equ:EDD_weighted_M} and \eqref{equ:EDDARL_weighted_M}, the performance of the weighted $M$ voting rule highly depends on the parameters $\bar{m}$ and $\bar{M}$ of critical syndromes. With the condition in \eqref{equ:condition_csynd_weighted}, these parameters can be obtained by solving the following 0-1 integer programming:
\begin{align}
\bar{m} = \mathop{\min \limits_{\omega \in \partial{M}}}~{\sum_{(l,k) \in \omega} {1}}\text{,  } 
\bar{M} = \mathop{\max \limits_{\omega \in \partial{M}}}~{\sum_{(l,k) \in \omega} {1}}\text{.}
\end{align}
The following fact describes bounds of $\bar{m}$ and $\bar{M}$, which will enable us to compute $\bar{m}$ and $\bar{M}$ with low complexity:

\begin{algorithm}
	\caption{Pruning Algorithm for $\bar{M}$}
	\begin{algorithmic}[1]
	\Require The weights for all sensors $\alpha_{l,k}$ and the threshold $M$
	\Ensure The maximum amount $\bar{M}$ of the elements for critical syndromes in $\partial{M}$ and its corresponding syndrome $\omega_{\bar{M}}$.
	    \State Set parent syndrome set $\Omega_{f} = \left\{ \{ (1,1), \cdots, (L,N_{L}) \} \right\}$, child syndrome set $\Omega_{c} = \emptyset$ and candidate set $\Omega_{cs}$.
        \State Set iteration index $i = N$ and $i_{min} = 1$.
	    \While {$i>i_{min}$}
        \State $i=i-1$.
        \State For each parent syndrome $\omega_{f} \in \Omega_{f}$ that contains sensors from group $l_1,l_2,\cdots,l_J$:
        \begin{itemize}
            \item Find $\bar{m}_{\omega_{f}}$ s.t. its least $\bar{m}_{\omega_{f}}$ weight sum meets \eqref{equ:condition_lowsum}. 
            \item Run the top-selection algorithm on $\omega_{f}$ to get $\omega_{f}^{ts}$ that satisfies \eqref{equ:condition_topsum}. Let $\Omega_{cs}$ obtain $\omega_{f}^{ts}$.
            \item Let $\omega_{fj} $ be the syndrome that drops the sensor of $\omega_{f}$ from group $l_j$ with the largest index, $1\leq j \leq J$. Include $\omega_{fj}$ into $\Omega_{c}$ if $\omega_{f}^{ts} \nsubseteq \omega_{fj}$ and the sum weight of $\omega_{fj}$ is greater than $M$.
        \end{itemize}
        \State Assign $\Omega_{f}$ to $\Omega_{c}$ and $i_{min}$ to $\mathop{\max \limits_{\omega_{f} \in \Omega_{f}}}~{\bar{m}_{\omega_{f}}}$.
		\EndWhile {$\textrm{ if }i \leq i_{min}$} or there exists $\omega_{f}^{ts}$ s.t. $\rvert \omega_{f}^{ts} \rvert = i$.
        \State Set $ \omega_{\bar{M}} = arg \mathop{\max \limits_{\omega \in \Omega_{cs}}}~{\rvert \omega \rvert}$, $\bar{M} = i \textrm{.}$
	\end{algorithmic}
	\label{alg:bar_M}
\end{algorithm}

\begin{fact} \label{fact:top_low_sum}
Let $\omega_{m}^{T}$ be the collection of the sensors with the top $m$ weights among all the network. If 
\begin{align}
{\sum_{(l,k) \in \omega_{m-1}^{T}}{\mathcal{I}_l}} < M\mathcal{I}_L \leq {\sum_{(l,k) \in \omega_{m}^{T}}{\mathcal{I}_l}} \textrm{,} \label{equ:condition_topsum}
\end{align}
then $\omega_{m}^{T} \in \partial M$. Moreover, $\bar{m} = m$. Also, let $\omega_{m}^{L}$ be the collection of the sensors with the $m$ least weights among all the network. If
\begin{align}
{\sum_{(l,k) \in \omega_{m-1}^{L}}{\mathcal{I}_l}} < M\mathcal{I}_L \leq {\sum_{(l,k) \in \omega_{m}^{L}}{\mathcal{I}_l}} \textrm{,} \label{equ:condition_lowsum}
\end{align}
then $\bar{M} \leq m$.
\end{fact}
\begin{IEEEproof}
    See Appendix~\ref{pf:top_low_sum}.
\end{IEEEproof}
First, $\bar{m}$ can be obtained quickly by a top-selection algorithm based on \eqref{equ:condition_topsum}. Moreover, to obtain $\bar{M}$, we leverage \eqref{equ:condition_lowsum} and the structure of the Hasse diagram to develop a pruning method for breadth-first search algorithm, as shown in Algorithm~\ref{alg:bar_M}. In particular, the complexity of the algorithm is limited by the maximum amount of possible critical syndromes in $\partial M$, as shown in the following fact.

\begin{fact} \label{fact:complexity_algo_pruning}
The total number of iterations in Algorithm~\ref{alg:bar_M} of parent syndromes is less than $C_{L-1}^{L-1+\ceil{n/2}}$.
\end{fact}
\begin{IEEEproof}
    See Appendix~\ref{pf:complexity_algo_pruning}.
\end{IEEEproof}

\end{remark}

\section{Simulation Results}\label{SimulationResults}

\begin{figure*}[tb] \vspace{-0.7cm}
\centering
\begin{tabular}{c c}
\subfloat[]{ \includegraphics[scale=0.47]{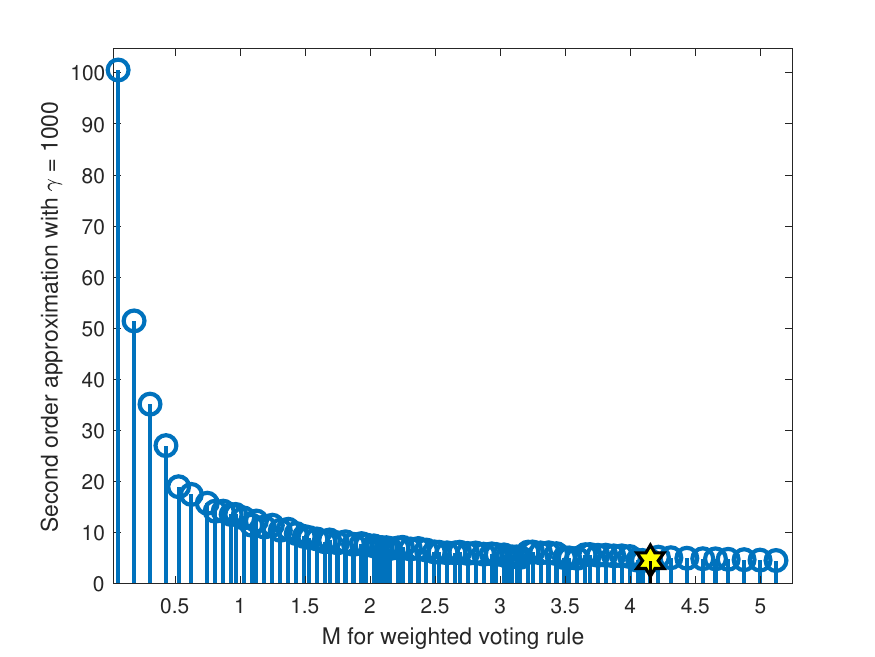} }
\subfloat[]{ \includegraphics[scale=0.47]{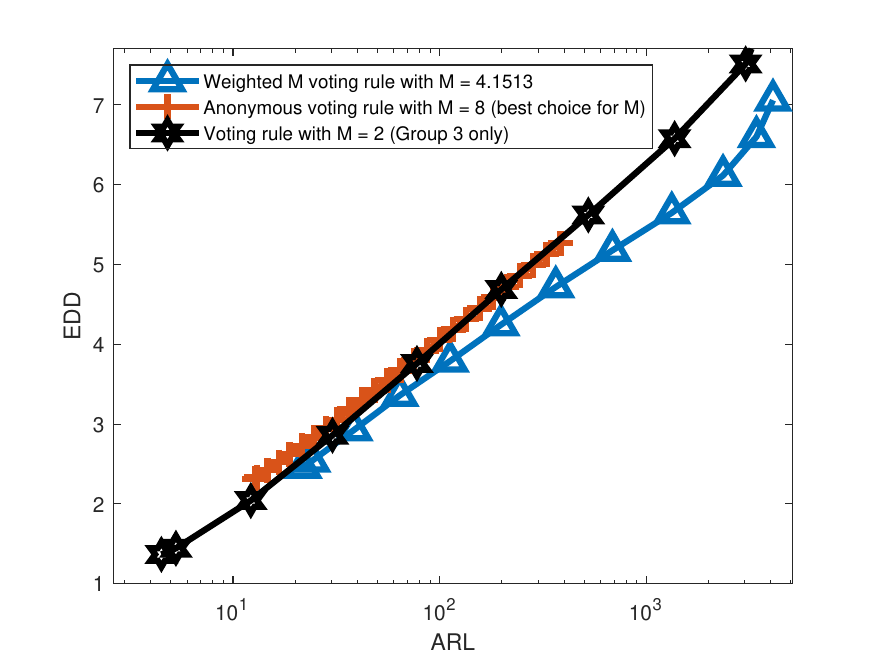} }
\end{tabular}\caption{(a) Second-order approximation of $E_0(\bar{\rho}_{M}( \mathbf{h},\boldsymbol{\alpha}))$ in case 1. (b) Comparison of $\bar{\rho}_{M}( \mathbf{h},\boldsymbol{\alpha})$, $\textrm{T}_{M}( \mathbf{h},\mathcal{G})$ and $\textrm{T}_{M}(\mathbf{h},\mathcal{G}_L)$ in case 1. The proposed $M$ for weighted $M$ voting rule is by (a).}
\label{fig:simu_hetnet}
\end{figure*}

\begin{figure*}[tb] \vspace{-0.6cm}
\centering
\begin{tabular}{c c}
\subfloat[]{ \includegraphics[scale=0.47]{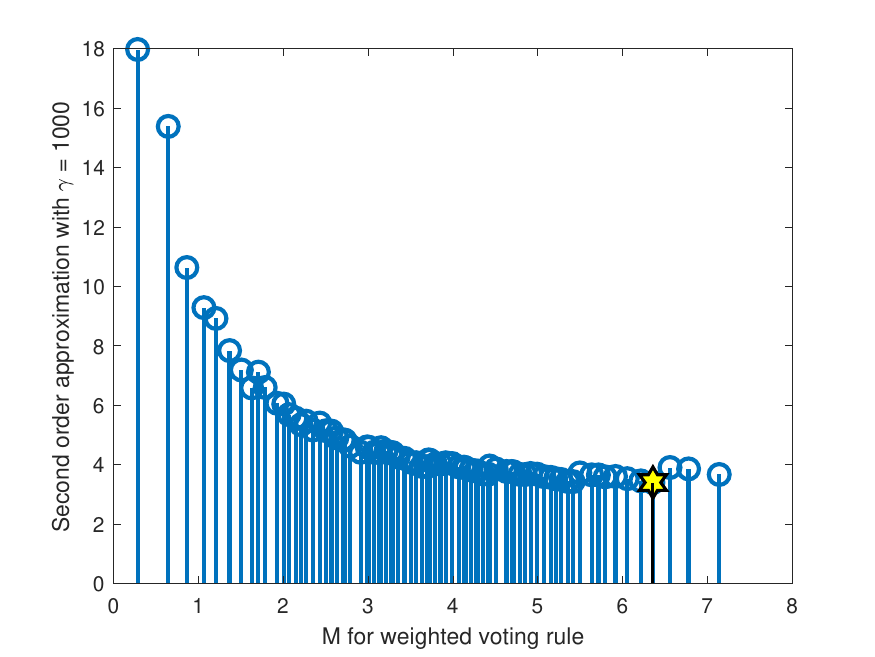} }
\subfloat[]{ \includegraphics[scale=0.47]{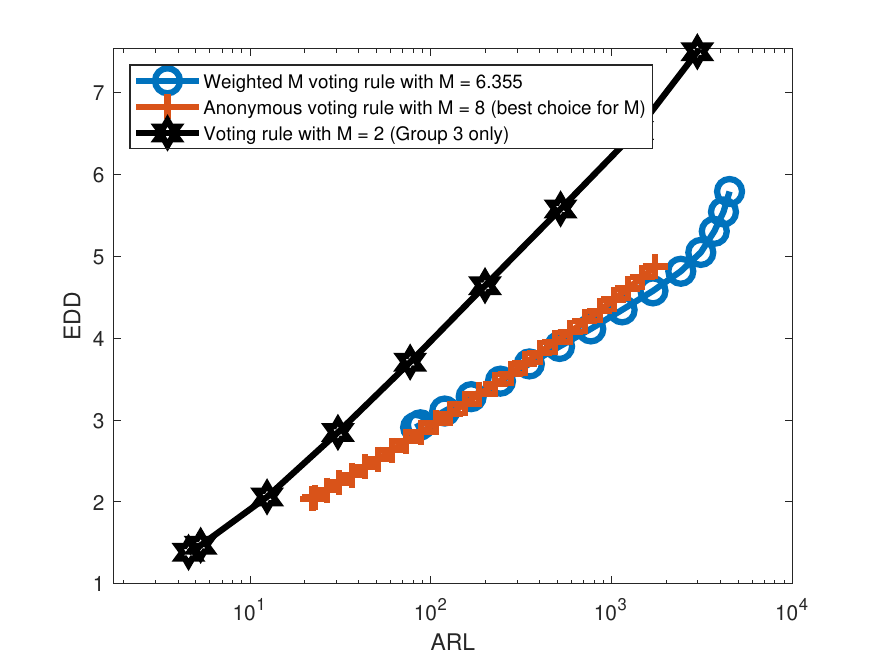} }
\end{tabular}\caption{(a) Second-order approximation of $E_0(\bar{\rho}_{M}( \mathbf{h},\boldsymbol{\alpha}))$ in case 2. (b) Comparison of $\bar{\rho}_{M}( \mathbf{h},\boldsymbol{\alpha})$, $\textrm{T}_{M}( \mathbf{h},\mathcal{G})$ and $\textrm{T}_{M}(\mathbf{h},\mathcal{G}_L)$ in case 2. The proposed $M$ for weighted $M$ voting rule is by (a).} \vspace{-0.2cm}
\label{fig:simu_homnet}
\end{figure*}

Our analytical results are verified by numerical simulations. We consider that there are three groups of sensors in total ($L=3$). The number of sensors in each group are $N_1 = 4$ and $N_{2} = N_{3} = 3$. The pre-change distributions of each sensor are assumed to be standard normal. To show the different heterogeneity, we set two cases of the post-change distributions:
\begin{itemize}
    \item Case 1 (High heterogeneity): $g_{l} \sim N(m_l,1)$, in which $(m_1, m_2, m_3) = (0.35, 0.75, 1)$.
    \item Case 2 (Low heterogeneity): $g_{l} \sim N(m_l,1)$, in which $(m_1, m_2, m_3) = (0.75, 0.85, 1)$.
\end{itemize}
For the weighted $M$ voting rule, the proposed $M$ is obtained by minimizing the second-order approximation \eqref{equ:EDDARL_weighted_M}, which requires parameters $\bar{M}$ and $\bar{\mathcal{D}}$. Among all possible $M$, we first determine $\bar{M}$ by Algorithm~\ref{alg:bar_M}, and then $\bar{\mathcal{D}}$ can be obtained by \eqref{equ:D_bar}. Fig.~\ref{fig:simu_hetnet}-(a) and Fig.~\ref{fig:simu_homnet}-(a) plot the second-order approximation versus $M$ when $\gamma$ in \eqref{equ:EDDARL_weighted_M} is set to $10^3$, where the yellow stars point out the suggested choice of $M$. Fig. \ref{fig:simu_hetnet}-(b) and Fig. \ref{fig:simu_homnet}-(b) further compare the performance between the weighted $M$ voting rule with the proposed $M$, the anonymous $M$ voting rule with the best choice of $M$ and the best case of the $M$ voting rule with the group $\mathcal{G}_3$. Both simulation results demonstrate that the weighted $M$ voting rule with the parameter $M$ suggested by our analysis steadly outperforms the other two fusion rules.

\section{Conclusions} \label{Conclusions}

In this work, we studied how to employ the knowledge of feedback identity in the HetDQCD problem with 1-bit feedback. In view of feedback syndrome, a general expression of simultaneous fusion rules had been discussed. Inspired by the Hasse diagram, the performance of these fusion rules can be derived via analyzing their critical syndromes. To concretely develop the syndrome-based fusion rule, we revisited the weighted $M$ voting rule, which only selects syndromes according to the weighted sum. A second-order approximation of weighted $M$ voting rule was obtained following the concept of syndromes. The threshold $M$ is then suggested from optimizing the second-order approximation, in which the coefficients are obtained by an efficient pruning algorithm that utilizes the properties of the Hasse diagram. The simulation results supported the weighted voting rule with proposed parameter setting, which is robust to both heterogeneous and homogeneous environments. Potential future works include investigating non-simultaneous fusion rules and the heterogeneity metric corresponding to the performance loss by anonymity.

\clearpage
\balance
\bibliographystyle{IEEEtran}
%\bibliography{reference}

\begin{thebibliography}{99}
\bibitem{ref:Page54}
E.~S. Page, ``{C}ontinuous inspection schemes,'' \emph{Biometrika}, vol.~41, no.~1, pp. 100--115, 1954.

\bibitem{ref:Lorden71}
G.~Lorden, ``{P}rocedures for reacting to a change in distribution,'' \emph{Ann. Math. Statist.}, vol.~42, no.~6, pp. 1897--1908, Dec. 1971.

\bibitem{ref:Moustakides86}
G.~V. Moustakides, ``{O}ptimal stopping times for detecting changes in distributions,'' \emph{Ann. Math. Statist.}, vol.~14, no.~4, pp. 1379--1387, Dec. 1971.

\bibitem{ref:Lai08}
L.~Lai, Y.~Fan, and H.~V. Poor, ``{Q}uickest detection in cognitive radio: {A} sequential change detection framework,'' in \emph{IEEE GLOBECOM 2008}, 2008, pp. 1--5.

\bibitem{ref:Lakhina04}
A.~Lakhina, M.~Crovella, and C.~Diot, ``{D}iagnosing network-wide traffic anomalies,'' \emph{ACM SIGCOMM Comput. Commun. Rev.}, vol.~34, no.~4, pp. 219--230, 2004.

\bibitem{ref:Nizam16}
F.~Nizam, S.~Chaki, S.~A. Mamun, and M.~S. Kaiser, ``{A}ttack detection and prevention in the cyber physical system,'' in \emph{Proc. Int. Conf. Comput. Commun. Informat. (ICCCI)}, 2016, pp. 1--6.

\bibitem{ref:Zhang23}
Q.~Zhang, Z.~Sun, L.~C. Herrera, and S.~Zou, ``{D}ata-driven quickest change detection in hidden markov models,'' in \emph{2023 IEEE International Symposium on Information Theory (ISIT)}, 2023, pp. 2643--2648.

\bibitem{ref:Liyan21_survey}
L.~Xie, S.~Zou, Y.~Xie, and V.~V. Veeravalli, ``Sequential (quickest) change detection: Classical results and new directions,'' \emph{IEEE Journal on Selected Areas in Information Theory}, vol.~2, no.~2, pp. 494--514, 2021.

\bibitem{ref:Mei05}
Y.~Mei, ``{I}nformation bounds and quickest change detection in decentralized decision systems,'' \emph{IEEE Trans. Inf. Theory}, vol.~51, no.~7, pp. 2669--2681, Jul. 2005.

\bibitem{ref:Banerjee16}
S.~Banerjee and G.~Fellouris, ``{D}ecentralized sequential change detection with ordered {CUSUM}s,'' in \emph{2016 IEEE International Symposium on Information Theory (ISIT)}, Jul. 2016, pp. 36--40.

\bibitem{ref:Fellouris18}
G.~Fellouris, E.~Bayraktar, and L.~Lai, ``Efficient {B}yzantine sequential change detection,'' \emph{IEEE Trans. Inf. Theory}, vol.~64, no.~5, p. 3346–3360, Jul. 2018.

\bibitem{ref:Huang21}
Y.-C. Huang, Y.-J. Huang, and S.-C. Lin, ``Asymptotic optimality in {B}yzantine distributed quickest change detection,'' \emph{IEEE Trans. Inf. Theory}, vol.~67, no.~9, pp. 5942--5962, Sep. 2021.

\bibitem{ref:Sun22}
Z.~Sun, S.~Zou, R.~Zhang, and Q.~Li, ``{Q}uickest change detection in anonymous heterogeneous sensor networks,'' \emph{IEEE Trans. Signal Processing}, vol.~70, pp. 1041--1055, 2022.

\bibitem{ref:Sun23}
Z.~Sun and S.~Zou, ``{Q}uickest anomaly detection in sensor networks with unlabeled samples,'' \emph{IEEE Trans. Signal Processing}, vol.~71, pp. 873--887, 2023.

\bibitem{ref:ISIT23}
W.-H. Li and Y.-C. Huang, ``{B}andwidth-constrained distributed quickest change detection in heterogeneous sensor networks: {A}nonymous vs non-anonymous settings,'' in \emph{2023 IEEE International Symposium on Information Theory (ISIT)}, 2023, pp. 1384--1389.

\bibitem{ref:Hasse}
J.~Gross, J.~Yellen, and M.~Anderson, \emph{Graph Theory and Its Applications}.\hskip 1em plus 0.5em minus 0.4em\relax CRC Press, 2018.


\end{thebibliography}

\clearpage

\appendix \label{Appendix}

\subsection{Proof of Equation \eqref{equ:2nd_ADD_mvote}} \label{pf:2nd_ADD_mvote}
\begin{IEEEproof}
Given the predetermined set $\mathcal{D}$ of selecting sensors, in \cite{ref:ISIT23} we defined another fusion rule $\textrm{T}^{(M)}$ called $M$-th alarm within $\mathcal{D}$ by
\begin{align*}
&\textrm{T}^{(M)}( \mathbf{h},\mathcal{D}) = \inf \left\{t: \lvert \{(l,k)\in\mathcal{D}:\mathop{\max \limits_{t}}~W_{t}^{l,k}>h_l \} \rvert \geq M \right\} \textrm{.}
\end{align*}
And in \cite[Theorem 3]{ref:ISIT23}, we proved that for $h \rightarrow \infty$,
\begin{align}
    &E_{0}[\textrm{T}^{(M)}( \mathbf{h},\mathcal{G})] = h + \xi _{M}\sqrt{h}(1+o(1)) \label{equ:appendix_EDD_malarm}\\
    &E_{0}[\textrm{T}_{M}( \mathbf{h},\mathcal{G})] \leq h + \xi _{M}\sqrt{h}(1+o(1))\textrm{.} \label{equ:appendix_EDD_mvote}
\end{align}
Therefore, equation \eqref{equ:2nd_ADD_mvote} holds by extending \eqref{equ:appendix_EDD_mvote} to equality, which follows from the fact that $\textrm{T}^{(M)}( \mathbf{h},\mathcal{G}) \leq \textrm{T}_{M}( \mathbf{h},\mathcal{G})$ and equation \eqref{equ:appendix_EDD_malarm}.
\end{IEEEproof}

\subsection{Proof of Theorem~\ref{thm:omega_fusion_rule}} \label{pf:omega_fusion_rule}
\begin{IEEEproof}
We first prove \eqref{equ:thres_omega}. Notice that for all $\omega \in \Omega$, the consensus rule on the subnetwork $\textrm{T}_{\rvert \omega \rvert}(\mathbf{h},\omega)$ always suffices to trigger the $\Omega$ fusion rule, leading to
\begin{align}
E_{\infty}(\textrm{T}_{\rvert \omega \rvert}(\mathbf{h},\omega)) \geq E_{\infty}(\rho_{\Omega}(\mathbf{h}) \textrm{.}
\end{align}
By setting  
\begin{align*}
\omega = arg\mathop{\min \limits_{\omega \in \partial \Omega}}~{\sum_{(l,k) \in \omega}{\mathcal{I}_l}}\textrm{,}
\end{align*}
according to \eqref{equ:FAR_mvote} we have
\begin{align*}
h_{\gamma,\Omega} \geq \frac {\log \gamma} {\mathcal{I}_{\Omega}}(1+o(1)). \label{equ:pf_thres_omega_1}
\end{align*}
Also, from the definition of critical syndrome in Definition ~\ref{def:critical_synd} it follows that
\begin{align}
    P_{\infty} (\rho_{\Omega}(\mathbf{h}) \leq z ) 
    &\leq 
    \sum_{\omega \in \partial \Omega}
    { \prod_{(l,k) \in \omega }{ P_{\infty} \left( W_{z}^{l,k} \geq \mathcal{I}_{l}h \right)}}\nonumber\\
    &\leq 
    \rvert \partial \Omega \rvert \max \limits_{\omega \in \partial \Omega} \prod_{(l,k) \in \omega}{ P_{\infty} \left( W_{z}^{l,k} \geq \mathcal{I}_{l}h \right)}\nonumber\\
    &\leq \rvert \partial \Omega \rvert \cdot e^{-\mathcal{I}_{\Omega}h} \equiv \gamma,
\end{align}
which implies
\begin{align}
    E_{\infty} [\rho_{\Omega}(\mathbf{h})] &= \int_{0}^{\infty}{P_{\infty} \left( \rho_{\Omega}(\mathbf{h}) \geq z  \right)dz}\nonumber\\
    &\geq \int_{0}^{1/\gamma}{P_{\infty} \left( \rho_{\Omega}(\mathbf{h}) \geq z \right)dz}\nonumber\\
    &\geq \int_{0}^{1/\gamma}{(1-\gamma)dz}=\frac{1}{\gamma}-1 \textrm{,}
\end{align}
and thus 
\begin{align}
h_{\gamma,\Omega} \leq \frac {\log \gamma} {\mathcal{I}_{\Omega}}(1+o(1)). \label{equ:pf_thres_omega_2}
\end{align}
Hence \eqref{equ:thres_omega} holds from \eqref{equ:pf_thres_omega_1} and \eqref{equ:pf_thres_omega_2}. It remains to prove \eqref{equ:EDD_omega}. Notice that $\textrm{T}_{\rvert \omega \rvert}(\mathbf{h},\omega)$ and the anonymous $m_{\Omega}$ voting rule are the necessary and sufficient conditions to trigger the $\Omega$ fusion rule, respectively. The bounds of \eqref{equ:EDD_omega} thus follow by applying \eqref{equ:2nd_ADD_mvote} to both fusion rules.
\end{IEEEproof}

\subsection{Proof of Theorem~\ref{thm:weighted_voting_rule}} \label{pf:weighted_voting_rule}
\begin{IEEEproof}
It is basically following Theorem~\ref{thm:omega_fusion_rule}. Notice that for any syndrome $\omega$ that triggers weighted $M$ voting rule, it follows that
\begin{align}
    {\sum_{(l,k) \in \omega}{\mathcal{I}_l}} \geq M\mathcal{I}_L \textrm{.}
\end{align}
Therefore \eqref{equ:thres_weighted_M} is obtained by replacing $\mathcal{I}_\Omega$ to $M\mathcal{I}_L$ in \eqref{equ:thres_omega}.

\end{IEEEproof}

\subsection{Proof of Fact~\ref{fact:top_low_sum}} \label{pf:top_low_sum}
\begin{IEEEproof}
For any syndrome $\omega$ such that $\rvert \omega \rvert = m$, it follows that
\begin{align}
    {\sum_{(l,k) \in \omega_{m}^{T}}{\mathcal{I}_l}} \geq {\sum_{(l,k) \in \omega}{\mathcal{I}_l}} \geq {\sum_{(l,k) \in \omega_{m}^{L}}{\mathcal{I}_l}} \textrm{.}
\end{align}
Therefore, if \eqref{equ:condition_topsum} holds, any collection with less than $m$ sensors cannot be the syndrome of weighted $M$ voting rule. Since \eqref{equ:condition_topsum} also implies that $\omega_{m}^{T}$ obeys the condition of critical syndrome \eqref{equ:condition_csynd_weighted}, this suffices to prove the first part of this fact. Similarly, if \eqref{equ:condition_lowsum} holds, every collection with not less than $m$ sensors must be the syndrome of weighted $M$ voting rule, which supports the second part of this fact.
\end{IEEEproof}

\subsection{Proof of Fact~\ref{fact:complexity_algo_pruning}} \label{pf:complexity_algo_pruning}
\begin{IEEEproof}
Notice that in Algorithm~\ref{alg:bar_M}, all parent syndromes will generate unique critical syndromes via the top-selection algorithm. For each child $\omega_{fj}$ of $\omega_f$, define by $n_{l_j}$ the amount of sensors chosen from group $l_j$. With the setting of algorithm, the composition of $\omega_{fj}$ must be
\begin{align*}
    \omega_{fj} = \left\{ (l_1,1),\cdots,(l_1,n_{l_1}),\cdots, (l_J,1),\cdots,(l_J,n_{l_J}) \right\} \textrm{.}
\end{align*}
Therefore the critical syndromes in $\Omega_{cs}$ must follow the above form, too. This implies that each critical syndrome obtained by this algorithm is fixed once the amount of sensors from each group is determined. Since all critical syndromes cannot contain each other, the case of maximal critical syndromes occurs if they all contain exactly $\ceil{n/2}$ sensors. Hence, the number of iterations in this algorithm is eventually bounded by the number of the $\ceil{n/2}$ combination with repetitions from $L$ groups, which is $C_{L-1}^{L-1+\ceil{n/2}}$.

\end{IEEEproof}

\end{document}